\documentstyle[psfig,prd,aps]{revtex}
\def \pom {{I\!\!P}}

\newcommand{\rk}{\mbox{\boldmath $k$}}
\newcommand{\rkn}{\mbox{$k$}}
\newcommand{\rqn}{\mbox{\boldmath $q$}}

\begin{document}
\draft
\title{The QCD Pomeron in Ultraperipheral Heavy Ion Collisions: \\ I. The Double $J/\Psi$ Production}
 \author{ V.P.
Gon\c{c}alves $^{1,\dag}$\footnotetext{$^{\dag}$ E-mail:
barros@ufpel.tche.br} and  M. V. T. Machado
$^{1,2,\star}$\footnotetext{$^{\star}$ E-mail: magnus@if.ufrgs.br} }
\address{$^1$ Instituto de F\'{\i}sica e Matem\'atica,  Universidade
Federal de Pelotas\\
Caixa Postal 354, CEP 96010-090, Pelotas, RS, BRAZIL}
\address{$^2$ High Energy Physics Phenomenology Group, IF-UFRGS\\
Caixa Postal 15051, CEP 91501-970, Porto Alegre, RS, BRAZIL}
 \maketitle

\begin{abstract}
The contribution of the QCD Pomeron to the process $AA \rightarrow
AA \, J/\Psi \,J/\Psi $ is discussed. We focus on the
photon-photon collision, with the quasi-real photon coming from
the Weizs\"acker - Williams spectrum of the  nuclei. We calculate
the cross section for this process considering the solution of the
LLA BFKL equation at zero momentum transfer using  a small $t$
approximation for the differential cross section of the
subprocess. Furthermore, the impact of non-leading corrections to
the BFKL equation is also analyzed. In both cases the cross
section is found to increase with the energy, predicting
considerable values for the LHC energies. Moreover, we compare our
results with the Born two-gluon approximation, which is energy
independent at the photon level. Our results indicate that the
experimental analyzes of this process can be useful to
discriminate the QCD dynamics at high energies.

\end{abstract}


\section{INTRODUCTION}

Understanding the behavior of high energy hadron reactions from a
fundamental perspective within QCD is an important goal of
particle physics. The behavior of scattering in the limit of high
energy and fixed momentum transfer is described in QCD, at least
in situations in which perturbation applies, by the
Balitskii-Fadin-Kuraev-Lipatov (BFKL) Pomeron \cite{bfkl}.
Attempts to test experimentally this sector of QCD have started in
last years, mainly considering  processes where specific
conditions are satisfied, which minimize the contributions of the
other mechanisms competing with the QCD Pomeron and guarantee the
validity  of the perturbative QCD methods.  In particular, the
virtualities of the gluons along the ladder should be large enough
to assure the applicability of the perturbative expansion. The
hard scale may be  provided either by the coupling of the ladder to
scattering particles, which contain a hard scale themselves, or by
 a large momentum transfer carried by the gluons. Furthermore, to
distinguish the BFKL from DGLAP evolution effects it is convenient
to focus on processes in which the scales on both ends of the
ladder are of comparable size. Some examples are the  measurements
of forward jets in deeply inelastic events at low values of the
Bjorken variable $x$ in lepton-hadron scattering, jet production
at large rapidity separations in hadron-hadron collisions and
off-shell photon scattering at high energy in $e^+\,e^-$
colliders, where the photons are produced from the leptons beams
by bremsstrahlung (For a recent review of BFKL searches, see e.g.
Ref. \cite{reviewhighenergy}). This last process presents some
theoretical advantages as a probe of QCD Pomeron dynamics compared
to the other ones because it does not involve a nonperturbative
target \cite{gamagama,gambrod,gamboone}. Moreover, such reaction
presents analogies with the process of scattering of two
quarkonia, which has been proposed as a gedanken experiment to
investigate the high energy regime in QCD \cite{muedip}. In such a
case, nonperturbative effects are suppressed by the smallness of
the quarkonium radius.

Other possibility for the study of the QCD Pomeron is the vector
meson pairs production in $\gamma \gamma$ collisions
\cite{ginzburg}. At very high energies $s \gg  -t$, diffractive
processes such as $\gamma \gamma \rightarrow $ neutral vector (or
pseudoscalar) meson pairs with virtual or real photons can test
the QCD Pomeron (Odderon) in a detailed way utilizing the simplest
possible initial state. As in the case of the large angle
exclusive $\gamma \gamma$ processes, the scattering amplitude is
computed by convoluting the hard scattering pQCD amplitude for
$\gamma \gamma \rightarrow q\overline{q} q\overline{q}$ with the
vector meson wave functions. For heavy vector mesons, this cross
section can be calculated using the perturbative QCD methods.
First calculations considering the Born two-gluon approximation
have been done in Refs. \cite{ginzburg}. More recently,
 the double $J/\Psi$ production in $\gamma \gamma$ collisions has been
proposed as a probe of the hard QCD Pomeron \cite{motyka}. There,  the hard QCD pomeron is
presumably the dominant mechanism. Theoretical estimates of the
cross sections presented in \cite{motyka} have demonstrated that
measurement of this reaction at a Photon Collider should be
feasible.

 In this paper we study the possibilities for investigating
QCD Pomeron effects in a different context, namely  in
photon-photon scattering at ultraperipheral heavy ion collisions.
In this case, the cross sections are enhanced since the $\gamma
\gamma$ luminosity increases as $Z^4$, where $Z$ is the atomic
number \cite{bert,reviewbaur}. Here, we will focus our analyzes on
double diffractive $J/\Psi$ production in $\gamma \gamma$
collisions, with the photons coming from the Weizs\"acker -
Williams spectrum of the nuclei. This process is unique since in
principle it allows to test the QCD Pomeron for arbitrary momentum
transfers, where  the hard scale is already provided by the
relative large mass of the charm quark. Here, our  goal  is
twofold:  to analyze the potentiality of this process to
constraint the QCD dynamics at high values of energy and to
provide reliable estimates for the cross sections concerning that
reaction. We calculate the cross section considering suitable
approximations  for the $\gamma \gamma \rightarrow J/\Psi\,J/\Psi$
subprocess. In particular, we estimate it  considering the
two-gluon exchange (Born level), the leading order solution of the
BFKL equation, as well as estimate the corrections on the energy
dependence  associated to the next-to-leading effects through the
Pomeron intercept (for a review on NLO BFKL corrections, see e.g.
Ref. \cite{nlosalam}  and references therein). Shortly, we have
found  that the two-gluon exchange mechanism gives rise to a
constant $\gamma \gamma$ cross section at large c.m.s. two-photon
energy $W$. To higher orders in perturbation theory, the iteration
of gluons in the $t$-channel promotes this constant to logarithms,
and the perturbative expansion of the cross section at high energy
has the form,
\begin{eqnarray}
\sigma_{\gamma \gamma} \propto \sigma^{(0)}\left[1 +
\sum_{k=1}^{\infty} a_k\,[\,\alpha_s \ln (W^2/s_0)\,]^k \right]\,+
\ldots \,,
\end{eqnarray}
where $s_0$ is the scale of the order of the charm mass squared,
the summation represents the series of the leading logarithms on
energy  to all orders in the strong coupling constant $\alpha_s$,
and the ellipsis stands for nonleading terms. The sum of the
leading logarithmic terms is made by the BFKL equation at leading
order, which predicts that cross section grows exponentially,
$\sigma \propto W^{2\omega_{\pom}}$, where $\omega_{\pom} = 4
\,\ln 2 \,N_c\,\alpha_s/\pi$ in the LLA solution. It has recently
been demonstrated that the NLO corrections to the BFKL equation
are large. The main effect is a reduced value of the so-called
Lipatov exponent $\omega_{\pom}$. Since then the effects of higher
orders have been studied  for measurable processes, but the
conclusions are not unambiguous \cite{reviewhighenergy}. This
situation should be improved in the future with the next
generation of linear colliders, such as the Japan Linear Collider
(JLC) at KEK, TeV Energy Superconducting Linear Accelerator
(TESLA) at DESY and CERN Linear Collider (CLIC), etc. However,
until these colliders become reality is important to consider
alternative searches in the current accelerators which allow us to
constrain the QCD dynamics.

This paper is organized as follows. In next section (Section
\ref{per}) we present a brief review of the ultraperipheral heavy
ion collisions and the main formulae to describe the photon-photon
process in these reactions. In Section \ref{sec3} we consider the
double $J/\Psi$ production in $\gamma \gamma$ collisions and
estimate this cross section considering distinct approximations to
the QCD Pomeron dynamics. There,  we calculate the total cross
sections for the subprocess for each approximation and   present
our results for double $J/\Psi$ production in ultraperipheral
heavy ion collisions. Finally, we present a summary of our main
conclusions in the Section \ref{conc}.

\section{Ultraperipheral Heavy Ion Collisions}
\label{per}

In ultraperipheral relativistic heavy-ion collisions the ions do
not interact directly with each other and move essentially
undisturbed along the beam direction \cite{bert}. The only
possible interaction is due to the long range electromagnetic
interaction and diffractive processes. Due to the coherent action
of all the protons in the nucleus, the electromagnetic field is
very strong and the resulting flux of equivalent photons is large.
A photon stemming from the electromagnetic field of one of the two
colliding nuclei can interact with one photon of the other nucleus
(two-photon process) or can penetrate into the other nucleus and
interact with its hadrons (photon-nucleus process). For a recent
review, see e.g. Ref. \cite{reviewbaur}. In particular, the
photoproduction of heavy quarks as a probe of the high density
effects have been recently emphasized in Refs. \cite{gelis}, where
the color glass condensate formalism \cite{mcl} was used to
estimate the cross section and transverse momentum spectrum.
Similarly, the elastic photoproduction of vector mesons in
ultraperipheral heavy ion collisions was studied recently in Ref.
\cite{vicber}, demonstrating that this process can be used to
constraint the nuclear gluon  distribution, which determines the
dynamics at high energies. For other studies of QCD dynamics in
one-photon processes see Ref. \cite{frank}. Here, we will restrict
our analyzes for the two-photon process and its potentiality to
investigate the QCD dynamics.

\begin{figure}[t]
\centerline{\psfig{file=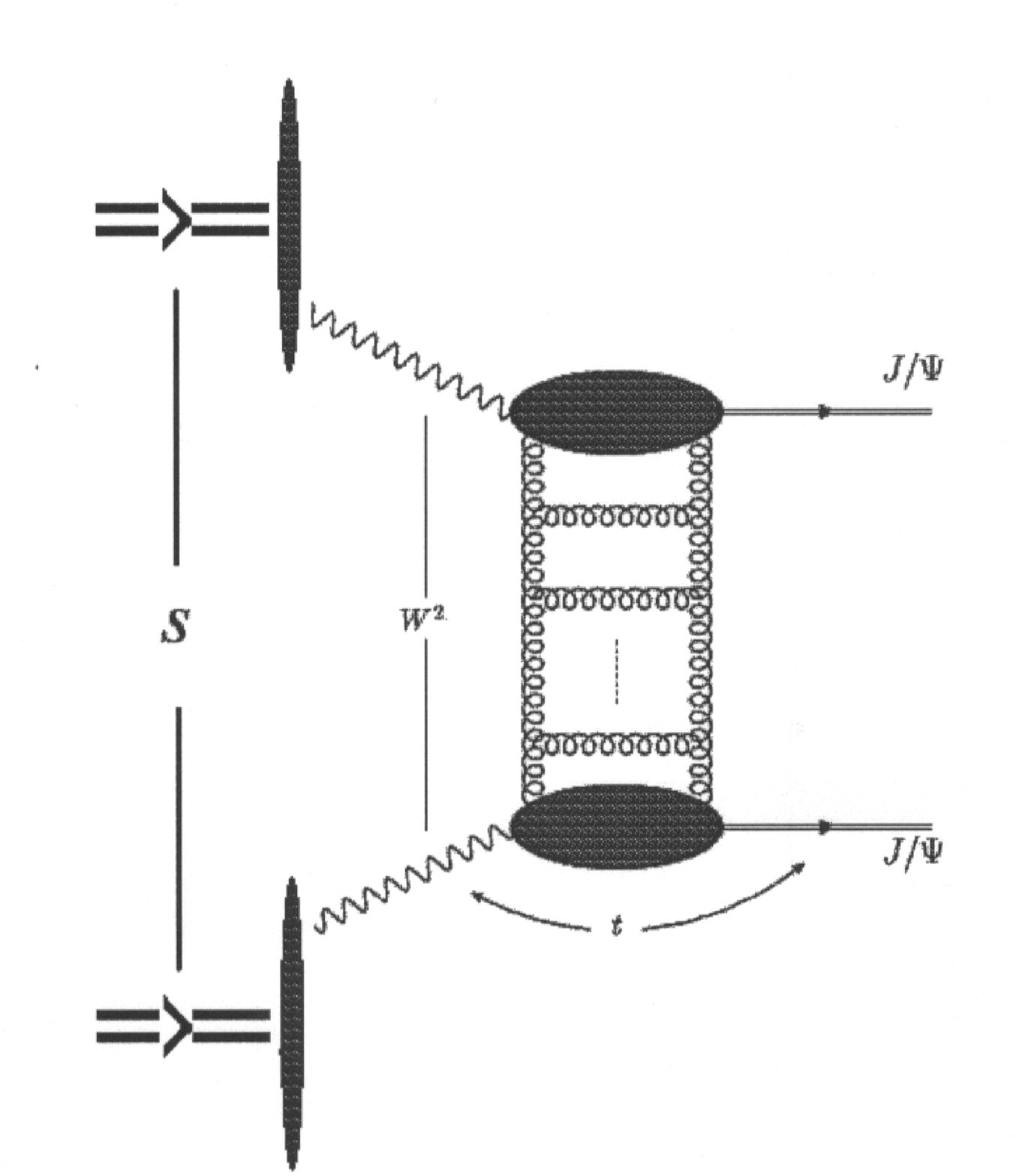,width=150mm}}
 \caption{The QCD pomeron exchange mechanism in ultraperipheral heavy ion collisions.}
\label{feynman}
\end{figure}

Relativistic heavy-ion collisions are a potentially prolific
source of $\gamma \gamma$ collisions at high energy colliders. The
advantage of using heavy ions is that the cross sections varies as
$Z^4 \alpha^4$ rather just as $\alpha^4$. Moreover, the maximum
$\gamma \gamma$ collision energy  $W_{\gamma \gamma}$ is $2\gamma
/R_A$,  about 6 GeV at RHIC and 200 GeV at LHC, where $R_A$ is the
nuclear radius and $\gamma$ is the center-of-mass system Lorentz
factor of each ion. In particular, the LHC will have a significant
energy and luminosity reach beyond LEP2, and could be a bridge to
$\gamma \gamma$ collisions at a future $e^+ e^-$ linear collider.
For two-photon collisions, the cross section for the reaction $AA
\rightarrow AA \,J/\Psi\, J/\Psi$, represented in Fig.
\ref{feynman}, will be given by
\begin{eqnarray}
\sigma_{AA \rightarrow AA \,J/\Psi \,J/\Psi} = \int \frac{d
\omega_1}{\omega_1} \, n_1(\omega_1) \int \frac{d
\omega_2}{\omega_2}\, n_2(\omega_2) \,\sigma_{\gamma \gamma
\rightarrow J/\Psi J/\Psi} (W = \sqrt{4 \omega_1 \omega_2}\,)
\,\,,
\end{eqnarray}
where the photon energy distribution $n(\omega)$ is calculated
within the equivalent photon or Fermi-Weizs\"acker-Williams (FWW)
approximation \cite{BN93}. In general, the  total cross section $
AA \rightarrow AA \,\gamma \gamma \rightarrow AA\, X$, where $X$ is
the system produced within the rapidity gap, factorizes into the
photon-photon luminosity $\frac{d{\cal{L}}_{\gamma
\gamma}}{d\tau}$ and the cross section of the $\gamma \gamma$
interaction,
\begin{eqnarray}
\sigma_{AA \rightarrow AA \,J/\Psi\, J/\Psi}(s) = \int d\tau \, \frac{d
{\cal{L}}_{\gamma \gamma}}{d\tau} \, \hat \sigma_{\gamma \gamma
\rightarrow J/\Psi\, J/\Psi}(\hat s), \label{sigfoton}
\end{eqnarray}
where $\tau = {\hat s}/s$, $\hat s = W^2$ is the square of the
center of mass (c.m.s.) system energy of the two photons and $s$
of the ion-ion system. The $\gamma \gamma$ luminosity  is given by
the convolution of the photon fluxes from two ultrarelativistic
nuclei:
\begin{eqnarray}
\frac{d\, {\cal{L}}_{\gamma \gamma}\,(\tau)}{d\tau} = \int ^1 _\tau
\frac{dx}{x} f(x)\, f(\tau/x),
\end{eqnarray}
where the photon distribution function $f(x)$ is related to the
equivalent photon number $n(\omega)$ via $f(x) =
(E/\omega)\,n(xE)$, with $x = \omega/E$ and $E$ is the total
energy of the initial particle in a given reference frame. The
remaining quantity to be determined in order to proceed  is the
quantity $f(x)$, which has been investigated by several groups
(for more details, see e.g. \cite{reviewbaur}). Here, we consider
the photon distribution of Ref.\cite{cahn}, providing a photon
distribution which is not factorizable. The authors of \cite{cahn}
produced practical parametric expressions  for the differential
luminosity by adjusting the theoretical results, which reads as,
\begin{equation}
\frac{d{\cal L}_{\gamma \gamma}}{d\tau}=\left(\frac{Z^2
\alpha}{\pi}\right)^2\, \frac{16}{3\tau}\, \, \,\times\,  \left\{ \begin{array}{cc}
 \, \mbox{$\xi (z)$}, & \mbox{$0.05<z<5$}  \\
\, \mbox{$\left[\ln{\left(\frac{1.234}{z}\right)}\right]^3$}, &  \mbox{$z<0.05$}
\end{array}
\right.
 \label{e3}
\end{equation}
where $z=mR/\gamma = 2MR\sqrt{\tau}$, $m$ is the mass of the
system produced in the two photon collision, $M$ is the nucleus
mass, $R$ its radius and $\xi(z)$ is given by,
\begin{equation}
\xi(z)=\sum_{i=1}^{3} A_{i} e^{-b_{i}z},
\label{e4}
\end{equation}
from an adjust to the numerical integration of the photon
distribution, with an accuracy of  $\sim 2\% $ in the referred
region on $z$. The adjustable parameters are the following:
$A_{1}=1.909$, $A_{2}=12.35$, $A_{3}=46.28$, $b_{1}=2.566$,
$b_{2}=4.948$, and $b_{3}=15.21$.

The approach given above excludes possible final state
interactions of the produced particles with the colliding
particles, allowing reliable calculations of ultraperipheral heavy
ion collisions. Therefore, to estimate the double $J/\Psi$
production it is only necessary to consider a suitable  QCD model
for the double heavy meson production. In lines of the analysis
presented here, it is worth mentioning that estimates for the
double pion (light mesons)  production in  ultraperipheral heavy
ion collisions have been presented in Ref. \cite{NataleRoldao}. In
addition, there it was found a negligible contribution of
pomeron-pomeron (considered in the Regge approach) interaction for
very  heavy ions, whereas this is non-negligible for lightest
ions.

\section{Double $J/\Psi$ Production}
\label{sec3}

The process  $\gamma \gamma \rightarrow J/\Psi J/\Psi$ is a clean
reaction testing the BFKL Pomeron physics and provides estimates
for the heavy mesons production in photon induced processes. The
hard scale involved in the reaction, that is the charm mass, and
the  presence of the photon as the initial state  turns out it
suitable for perturbative treatment. The non-perturbative content
is provided only by the $J/\Psi$ light-cone wave function, which
is well constrained through the experimental measurement of its
leptonic width $\Gamma_{J/\Psi\rightarrow \, l^+ l^-}$. In the
following we use the high energy factorization and the BFKL
dynamics in order to perform estimates for the referred reaction.
Our  general formulae for the differential cross section and
imaginary part of the scattering amplitude are given as follows
\cite{motyka},

\begin{eqnarray}
\frac{d\sigma\,(\gamma \gamma \rightarrow  J/\Psi \, J/\Psi)}{dt} & = & \frac{|{\cal A}(W^2,t) |^2}{16\,\pi} \,,\label{dsigdt}\\
{\cal I}m\,{\cal A}(W^2,t) & = &  \int \frac{d^2\rk}{\pi} \,
\frac{\Phi_{\gamma \, J/\Psi}(\rk,\rqn) \,\,
\widetilde{\Phi}_{\gamma \, J/\Psi}(W^2,\rk,\rqn)}{(\rk + \rqn/2
)^2 \,\,\, (\rk - \rqn/2  )^2}\,,
\end{eqnarray}
where $W$ is the center of mass energy of the two photon system
and the  photon-meson  impact factor is denoted by $\Phi_{\gamma
\, J/\Psi}$. At the Born level $\widetilde{\Phi} = \Phi$ and the
reaction is described by the two-gluon (a bare approximation for
the QCD Pomeron), which have transverse momenta $ \rqn/2 \pm \rk$
and where the momentum transfer is $t=-\rqn^2$. When considering
the complete gluon ladder contribution, the quantity
$\widetilde{\Phi}$ contains the impact factor and the gluon
emission on the ladder, which is driven by the QCD dynamics. At
the LLA level, the BFKL ladder contribution for the $t$-channel
exchange provides the following expression for it,

\begin{eqnarray}
\widetilde{\Phi}_{\gamma \, J/\Psi}(W^2,\rk,\rqn) & = & \int
d^2\rk^{\prime} \,\frac{\rk^2}{\rk^{\prime\,2}} \,
F(W^2,\rk,\rk^{\prime},\rqn)\,  \Phi_{\gamma \,
J/\Psi}(\rk^{\prime},\rqn) \,\,,\label{IPBFKL}
\end{eqnarray}
where
\begin{eqnarray}
 \Phi_{\gamma \, J/\Psi}(\rk,\rqn) & = & {\cal C} \, \left[ \frac{1}{m^2_c + \rqn^2} -
\frac{1}{m^2_c+ \rk^2} \right]\,\,, \label{IP}
\end{eqnarray}
and $F(W^2,\rk,\rk^{\prime},\rqn)$ is the solution of the LLA BFKL
equation at $t\neq 0$. The Eq. (\ref{IP}) defines the impact
factor in the nonrelativistic approximation. We have considered
the following parameters for the further calculations:
$m_c=m_{J/\Psi}/2=1.55$ GeV, ${\cal C}= \sqrt{\alpha_{em}}
\,\alpha_s(\mu^2)\, e_c\, \frac{8}{3} \,\pi \, m_c \,f_{J/\Psi}$,
with  $e_c=2/3$ and $ f_{J/\Psi}=0.38$ GeV.

\begin{figure}[t]
\centerline{\psfig{file=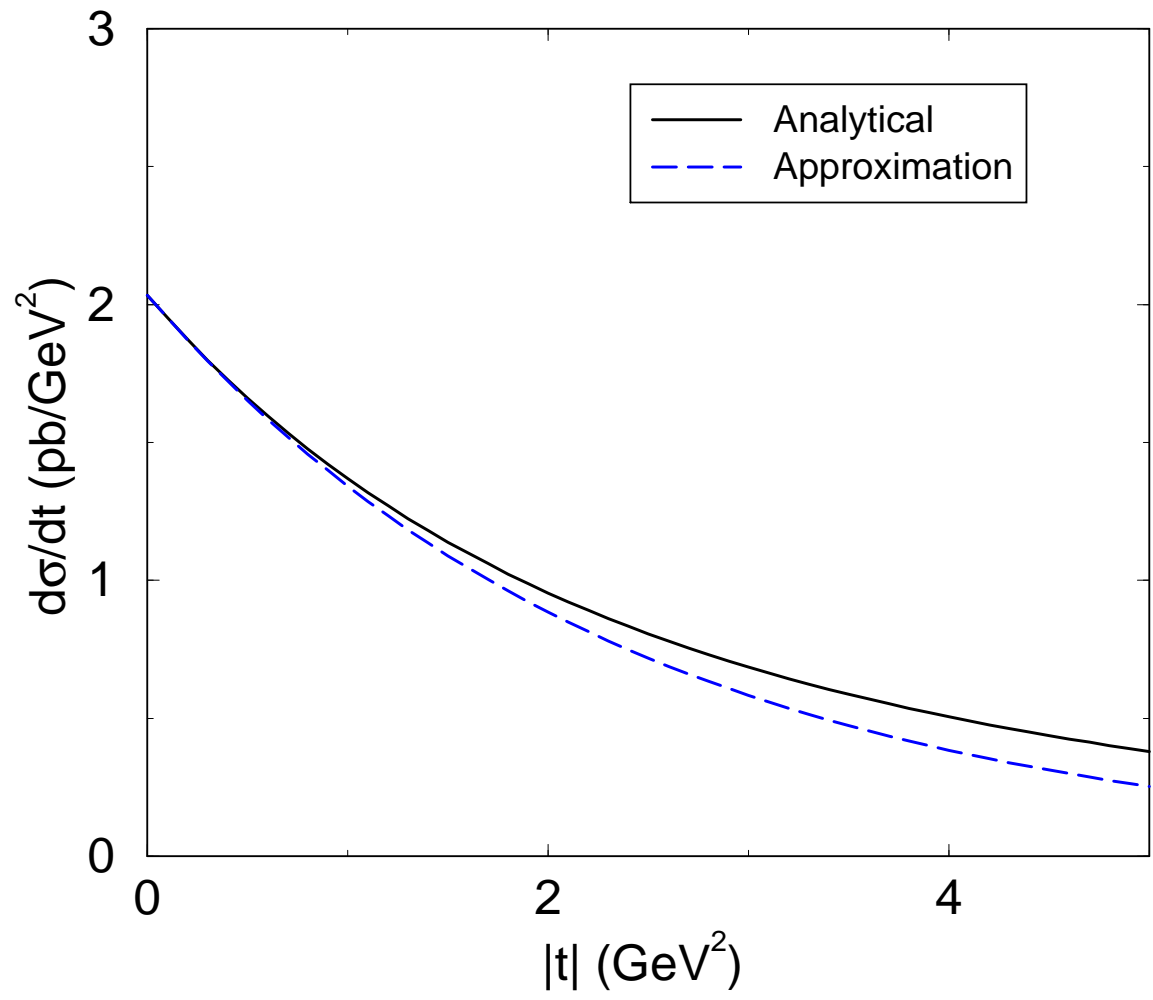,width=150mm}}
 \caption{ The comparison between the Born two-gluon analytical calculation
 (solid line) for the differential cross section and the small-$t$ approximation (dashed line). }
\label{dsdtb}
\end{figure}

In order to compute the total cross section for the process, we
would need consider the LLA BFKL non-forward solution
\cite{Lipatov86} (for a derivation in the momentum space, see e.g.
\cite{Forshaw97}) in Eq. (\ref{IPBFKL}), put all together to
obtain the amplitude  and then integrate Eq. (\ref{dsigdt}) over
$0\leq |t|\leq \infty$. Further refinements can be considered. For
instance, the infrared contributions by modifying the gluon
propagator can be still  investigated, as done in Ref.
\cite{PLBWerner} for the Born case and as in Ref. \cite{motyka}
for the full resummation.  Moreover, non-leading corrections could
be also implemented in a numerical calculation, for instance in
lines of Ref. \cite{motyka}.   Here, we follow a different
procedure, since we are interested mostly in the energy dependence
of the process. We will consider a small $t$ approximation,
providing accuracy enough for our purposes in the present work, in
a such way that the total cross section can be written as,
\begin{eqnarray}
\sigma_{tot} \, (\gamma \,\gamma \rightarrow J/\Psi\, J/\Psi) & =
& \frac{1}{B_{J/\Psi\,J/\Psi}}\,\left. \frac{d\sigma \,(\gamma
\,\gamma \rightarrow J/\Psi\, J/\Psi)}{dt}\,\right|_{t=0}\,,
\label{smalltapprox}
\end{eqnarray}
with
\begin{eqnarray}
 \frac{d\sigma \,(\gamma \,\gamma \rightarrow
J/\Psi\, J/\Psi)}{dt} & = &  \frac{|{\cal A}(W^2,t=0)|^2}{16\,\pi}
\, \exp \left(-B_{J/\Psi \, J/\Psi}\cdot |t| \right)\,,
\label{dsdtap}
\end{eqnarray}
where $B_{J/\Psi\, J/\Psi}$ is the corresponding slope parameter.
This quantity is not available from experimental measurements,
instead of the slope $B_{J/\Psi\,p}$ obtained from $J/\Psi$
photoproduction \cite{jpsiphotop}. In the following we should
estimate  this quantity using information about the Born two-gluon
approximation, where $F=\delta^{2}(\rk - \rk^{\prime})$ . In this
particular case the amplitude reads as,
\begin{eqnarray}
{\cal A}_{Born}(W^2,t)  =   \int \frac{d^2\rk}{\pi} \,
\frac{\Phi_{\gamma \, J/\Psi}(\rk,\rqn) \,\,\Phi_{\gamma \,
J/\Psi}(\rk,\rqn)}{(\rk + \rqn/2 )^2 \,\,\, (\rk - \rqn/2  )^2}\,,
\end{eqnarray}
which can be computed analytically if one considers a  strong
coupling constant not depending on $\rkn^2$. Using the impact
factor given by Eq. (\ref{IP}) and performing the integration over
gluon the transverse momentum, one obtains
\begin{eqnarray}
{\cal A}_{Born}(W^2,t) = \frac{16\,{\cal C}^2}{m_c^2\,(4\,m_c^2 +
|t|\,)^2}\,. \label{Bornggt}
\end{eqnarray}
Now, we are able to use Eq. (\ref{Bornggt}) in order to estimate the $B$ parameter
for the double $J/\Psi$ production. This quantity can be computed from the scattering
amplitude above by the small $t$ approximation,
\begin{eqnarray}
B_{J/\Psi\, J/\Psi} & = &  \lim_{|t| \rightarrow 0} \,
\frac{d}{dt}\,\left[ \ln\,\left( \frac{d\sigma_{Born}}{dt}\right)
\right] = \lim_{|t| \rightarrow 0} \, \frac{d}{dt}\,\left[
\ln\,\left( \frac{16\,{\cal C}^4}{\pi\,m_c^4\,(4\,m_c^2 +
|t|\,)^4}\right) \right]\,,
 \label{dsigdtanalyt}
\end{eqnarray}
which gives a slope parameter with value  $B_{J/\Psi\,
J/\Psi}=m_c^{-2}$. Moreover, we expect that this estimate has
little sensitivity on  energy, since it is known that the hard
Pomeron gives a low $\alpha^{\prime}_{\pom}$ (almost no shrinkage)
\cite{Predazzi02}. That is, the phenomenology on $J/\Psi$
photoproduction has provided $\alpha^{\prime}_{\pom}\simeq 0.1$
\cite{jpsiphotop}, instead of the usual $\alpha^{\prime}_{\pom}=
0.25$ from the hadronic analysis, where the corresponding slope is
given by $B = b_0 + 2\,\alpha^{\prime}_{\pom}\,\ln\,(s/s_0)$ in
the Regge analysis. We have verified also the accuracy of our
approximation comparing the analytical calculation, Eq.
(\ref{dsigdtanalyt}), with the small-$t$ estimate given by Eq.
(\ref{dsdtap}). The comparison is shown in Fig. \ref{dsdtb}, for a
fixed coupling constant $\alpha_s(m_{J/\Psi}^2)\simeq 0.289$, in
agreement with the estimates at Ref. \cite{motyka}. It is worth
mentioning that the typical values measured in elastic $J/\Psi$
photoproduction stay in the range $|t|<1.5$ GeV$^{-2}$, where our
approximation produces the same result as the analytical case.

Having the slope parameter $B$, and taking  Eq.
(\ref{smalltapprox}),  one can estimate the total cross section
considering the Born two-gluon exchange. It is simple to verify
that  the normalization for $\sigma_{tot}$ depends  significantly
on the strong coupling $\alpha_s$. We have obtained that the cross
section takes values  of order a few pb's, having  a lower bound
if  the typical HERA value $\alpha_s=0.2$ is used as well as  an
upper bound for $\alpha_s(m^2_c)$. These estimates are
corroborated by previous calculations \cite{ginzburg}, and some
results are presented in the Table \ref{tabela}, considering
different values of $\alpha_s$. The results are also consistent
with  Ref. \cite{motyka}, where the calculations were done beyond
the leading log approximation. For instance, we have found
$\sigma_{tot}=3.6$ pb using the running $\alpha_s(\rk^2 + m_c^2)$
and $s_0=0.16 \, GeV^2$ (this is the parameter accounting for the
infrared cutoff). The refined study of \cite{motyka} gives a close
result, $\sigma_{tot} \approx 2-2.6$ pb. Our estimates are quite
reliable in view of the approximations considered here. It should
be stressed that other additional sources of uncertainty are the
charm mass and the nonrelativistic approximation  for the impact
factor.

In order to perform a LLA BFKL calculation, the following solution
for the evolution equation in the forward case was considered (see
the pedagogical reviews on Refs. \cite{Forshaw97,Predazzi02}),
\begin{eqnarray}
F(W^2,\rk, \rk^{\prime}) & = & \frac{1}{\sqrt{2\,\pi^3\, a \,\rk^2 \,\rk^{\prime\,2}}}\, %%@
\frac{1}{\sqrt{\ln (W^2/\tilde{s})}} \,\left( \frac{W^2}{\tilde{s}} \right)^{\omega_{\pom}} \, \exp %%@
\left[-\frac{\ln^2 (\rk^2/\rk^{\prime\,2})}{2\,a\,\ln (W^2/\tilde{s})}  \right] \,\,,
\end{eqnarray}
where
\begin{eqnarray}
 \omega_{\pom} & = & \frac{3\,\alpha_s}{\pi}
\, 4\, \ln \,2, \hspace{1cm} a=  \frac{3\,\alpha_s}{\pi}
\,28\,\zeta (3)
\end{eqnarray}
and $1+\omega_{\pom}$ is the Pomeron intercept in the leading
logarithmic approximation,  which depends on $\alpha_s$. We have
taken $\tilde{s}=1$ GeV$^2$. The results are sensitive to the
choice for the intercept, providing an enhancement of the total
cross section by one/two
 orders of magnitude in relation to the Born level at the
 considered energy range. For the further
studies we have  selected  the results using
$\alpha_s(\mu^2)=0.22$, typical at the HERA kinematical region.

\begin{figure}[t]
\centerline{\psfig{file=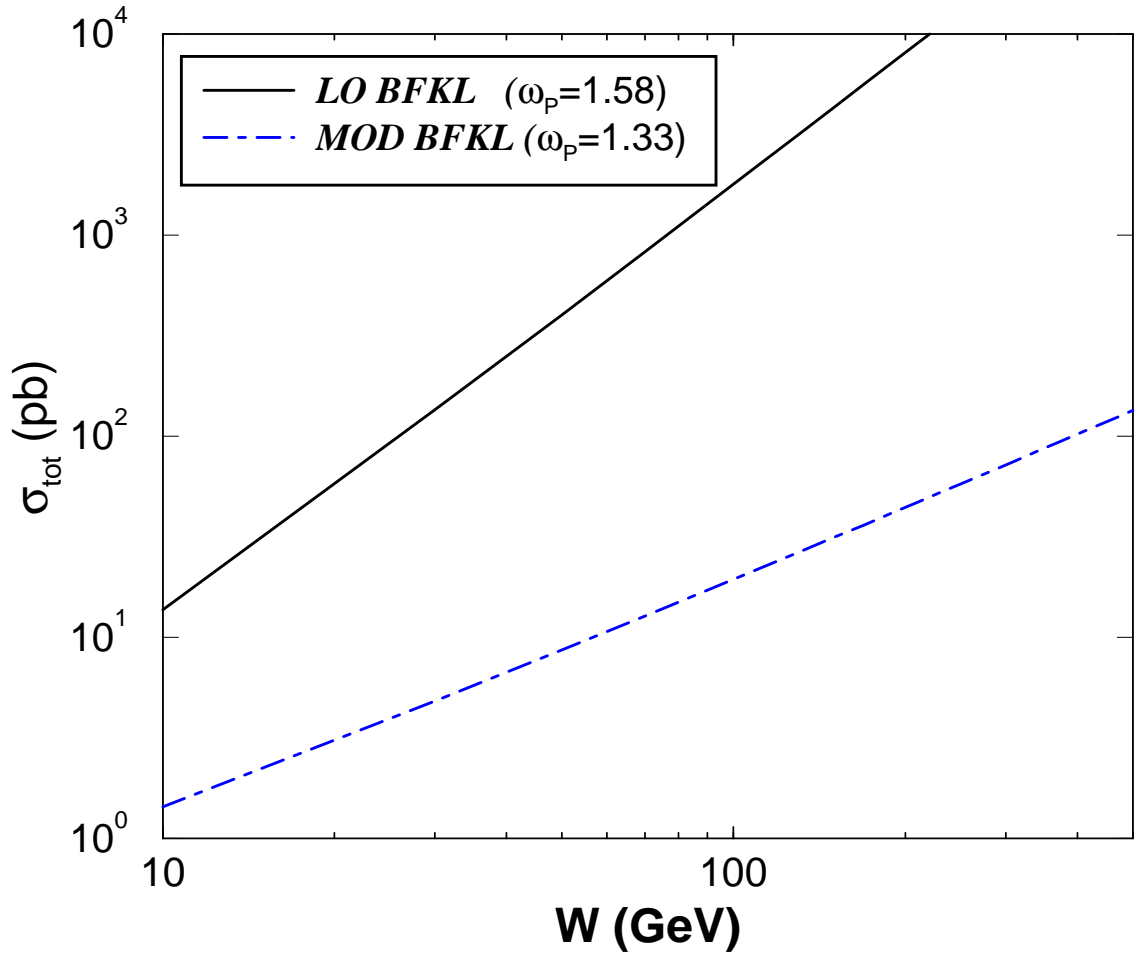,width=150mm}} \caption{ The
estimates for the total $\gamma \gamma$ cross section using LLA
BFKL solution (solid curve)  and non-leading effects (dot-dashed
curve). } \label{sigtot}
\end{figure}

The total cross section is shown in Fig. \ref{sigtot}, represented
by the solid line, producing an effective behavior given by
$W^{4\lambda}$ and where $\lambda=0.54$. The range on $W$ is the
region  possibly to be available at LEP2 and two-photon process in
ultraperipheral heavy ions collisions at LHC. The behavior is
considerably steep on energy and it is timely ask by the NLO
corrections to the LLA BFKL equation or unitarity corrections to
the LO calculation. Concerning the NLO effects, the correction
looks like
$\omega_{\pom}^{NLO}=\omega_{\pom}\,(1-N\,\overline{\alpha}_s)$,
where $\overline{\alpha}_s=3\,\alpha_s/\pi$. The value $N=6.5$,
obtained in the literature \cite{Lipatovnlo}, turns out the
corrections too large and even producing negative values if large
$\alpha_s$ values are considered. The complete resummation of the
non-leading effects, mostly collinear-enhanced contributions,  is
currently argued in order to provide reliable calculations (see
Ref. \cite{nlosalam} for a review).

For practical purpose in our further investigation below, we have
considered the modification of the Pomeron intercept in order to
be consistent with the stable higher order resummation results
\cite{NLOnew}, which give $\omega_{\pom}\simeq 0.3$ at the HERA
kinematic regime. Even a lower effective exponent is found
concerning the NLO BFKL using the BLM scheme for the
renormalization scale setting \cite{NLOBLM}. For a full
calculation we would need the  NLO impact factor, which is not
completely known \cite{bartels}. However, one can use the LO
impact factor, assuming that the main energy-dependent NLO
corrections come from the NLO BFKL sub-process rather than from
the photon impact factors \cite{NLOBLM}. In order to simulate the
NLO effects, we have used a value $\omega_{\pom}=0.37$, shown in
the dot-dashed curve of Fig. \ref{sigtot}. The effective behavior
obtained is given by $W^{4\lambda}$, with $\lambda=0.29$, which
value is somewhat close to $\lambda \simeq 0.23-0.29$ obtained in
\cite{motyka}.

It would be timely compare our estimates with the more refined
works in \cite{motyka} and in Ref. \cite{donnachiedosh}. In the
former, the most part of the NLO correction comes from considering
a kinematical constraint.   That is, the effects from restricting
the range for the term of real emission in the BFKL equation,
imposed upon the available phase space, corresponding to the
requirement that the virtuality of exchanged gluons in the BFKL
ladder is dominated by the transverse momentum squared. Such a
procedure covers  up to 70 \% of the NLO corrections to the
Pomeron intercept. As discussed above, our results are similar. In
Ref. \cite{donnachiedosh}, the dipole-dipole approach is
considered and predictions to double meson (light and heavy)
production are performed. There, the result for the total cross
section $\gamma \gamma \rightarrow J/\Psi\, J/\Psi$ is one or two
order of magnitude below the presented here and in \cite{motyka}.

\begin{figure}[t]
\centerline{\psfig{file=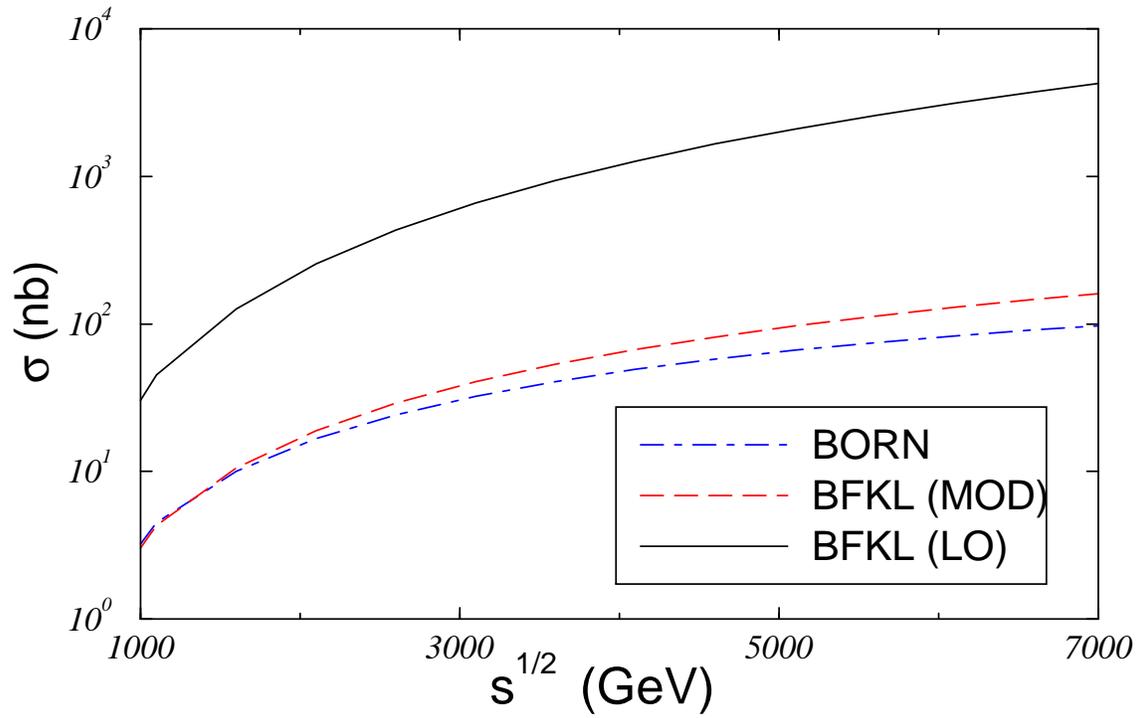,width=150mm}} \caption{The
estimates for the double $J/\Psi$ production in ultraperipheral
heavy ion collisions considering the LLA BFKL solution (solid
curve), non-leading effects (dashed curve) and the Born
approximation (dot-dashed curve).  } \label{sigtot2}
\end{figure}

Having determined the $\gamma \gamma$ cross section we can
estimate the double $J/\Psi$ cross section in ultraperipheral
heavy ion collision [Eq. (\ref{sigfoton})]. In Fig. \ref{sigtot2}
we present the energy dependence of the cross section, considering
the Pb + Pb collisions and distinct approximations for the $\gamma
\gamma$ subprocess. We have found that the LLA BFKL solution
predicts large values of cross section for LHC energies in
comparison with the Born approximation. Basically, we have that
the ratio between the BFKL prediction and the two-gluon cross
section is around 40 at $\sqrt{s} = 5500$ GeV. If the non-leading
corrections are considered, as discussed above, we have that this
ratio is reduced to approximately 1.7. Therefore, our results
indicate that if the NLO corrections to the BFKL approach are
account in the approximation considered here, a future
experimental analyzes of double $J/\Psi$ production in
ultraperipheral heavy ion collisions could not constrain the QCD
dynamics, since the BFKL(NLO) result is similar to the Born
prediction. Only accurate measurements would allow discriminate
between the two cases. Moreover, it is important to salient that
the NLO corrections can be larger, implying a full suppression of
energy enhancement associated with the iteration of gluons in the
$t$-channel  present in the QCD dynamics at high energies.
However, if the energy dependence of the $\gamma \gamma$ cross
section was driven for a large intercept, closer to the LO
prediction, the analyzes of ultraperipheral heavy ion collisions
can be useful.

Using the results for the cross section shown in Fig.
\ref{sigtot2} we can estimate the expected number of events for
the LHC luminosity. For PbPb collisions with energies of center of
mass equal to $\sqrt{s}= 5500 \, A\, GeV$, luminosities of
${\cal{L}}_{AA} = 4.2 \times 10^{26} \, cm^{-2}\,s^{-1}$ are used.
Consequently, during a standard $10^6\,s/$ month heavy ion run at
the LHC, we predict approximately  30, 50 and 1100 events for
Born, BFKL (MOD) and BFKL (LO), respectively. It is interesting to
compare these predictions with the results for double $J/\Psi$
production at LEP2, since the $\gamma \gamma$ center of mass
energies are similar. Considering an integrated luminosity of $500
\, pb^{-1}$ in  three years and $\sqrt{s}= 175\,GeV$, almost 70
events are expected taking into account the non-leading
corrections to the BFKL approach \cite{motyka}. Therefore, we
predict a large number of events in ultraperipheral heavy ion
collisions, allowing future experimental analyzes, even if the
acceptance for the $J/\Psi$ detection being low. It should be
stressed that the  LHC probably will operate  in its heavy ion
mode only four weeks per year.

Some comments related to background processes are in order here.
We did not consider the additional contribution of the
pomeron-pomeron process in the
 present calculations. It has been verified that such reactions
 would be non-negligible for light ions, while they are
 significantly suppressed for heavy ions \cite{NataleRoldao}. In
 the particular case of the double heavy meson production this
 contribution deserves more detailed studies, since the current
 treatments rely on the Regge formalism instead of a QCD approach.
 An important background for the photon-photon processes are the photonuclear
 interactions,  since the reactions have similar kinematics.  For the process consider here,
the diffractive $J/\Psi$ production in
 photon-pomeron interactions \cite{vicber,kleinvec} should contribute
 significantly. In particular, because the cross section for this process is large, the probability of having 
double (independent)  production of $J/\Psi$ in a  single nucleus-nucleus collision, associated to multiple interactions,  is non-negligible  \cite{kleinvec}.
However, in principle, an
 analyzes of the impact parameter dependence should allow to separate between the two
 classes of reactions, since two-photon interactions can occur at
 a significant distance from both nuclei, while a photonuclear
 interaction must occur inside or very  near a nucleus. 
We salient
 that the experimental separation between the two classes of
 processes is an important point, which deserves more studies.
 An additional contribution in two photon ultraperipheral collisions is
 the meson production accompanied by mutual Coulomb dissociation.
 It has been estimated that these reactions increases the total cross section
 for an amount of   $\sim 10\%$ at RHIC/LHC energies
 \cite{klein}. We disregarded this contribution in our
 calculations, since we believe that it is smaller than the
 theoretical uncertainty, associated, for example, to the choice of
 the running coupling constant, charm mass, etc.

\section{Summary}
\label{conc}

 Dedicated measurements have been proposed for
detecting and studying the large $1/x$ logarithm ressumation
effects in QCD. Experimentally establishing the BFKL effect in
data is very important for the understanding of the high energy
limit in QCD scattering. In the last years, many studies with this
objective have been realized,  but the conclusions are not
unambiguous. This situation should be improved in the future with
the next generation of linear colliders. As an alternative, in
this paper we  propose to investigate QCD Pomeron effects in a
different context, namely, in photon-photon scattering at
ultraperipheral heavy ion collisions.  We have  analyzed the
potentiality of this process to constraint the QCD dynamics at
high values of energy and provide reliable estimates for the cross
sections concerning that reaction. In particular, we analyze in
detail the double $J/\Psi$ production, considering distinct
approximations for the QCD dynamics. We have that the LO BFKL
cross section is much larger than the two-gluon cross section.
Unfortunately, the higher order corrections of the BFKL equation
are large, reducing the ratio between BFKL and two-gluon cross
section. It makes difficult to discriminate between non-leading
effects and the Born approximation. However, the number of events
predicted for LHC is large, allowing a future experimental
analyzes of this process as well as  more detailed studies about
the QCD dynamics.

As a last comment, it is important to salient that there are
limitations on the perturbative treatment that are intrinsic to
the BFKL equation, which come from the region of very high $s$.
The BFKL equation is  known to give rise to violation of the
unitarity bound at asymptotically large energies. Consequently,
the growth of the cross section predicted by the BFKL equation
cannot continue indefinitely, and unitarity corrections
(saturation effects) must arise to slow it down. They correspond
to multiple bare pomeron exchanges and multi-pomeron interactions,
taming the steep increase at high energies. We expect that the
inclusion of these effects in our calculations should imply in a
similar behavior of the non-leading BFKL predictions presented
here. More detailed studies related to the unitarity corrections
and BFKL dynamics will be analyzed in the sequel of this paper
\cite{vicmag2}, where we calculate the heavy quark production in
ultraperipheral heavy ion collisions and investigate the
saturation effects in the cross section.

\section*{ACKNOWLEDGMENTS}
The authors thank C. A. Bertulani, S. Klein, A. Natale and C. Rold\~ao for
helpful discussions.  We are particularly grateful to L. Motyka
for his careful reading of the manuscript and for his valuable comments. M.V.T.M. thanks the support of the High Energy Physics Phenomenology Group at Physics Institute, UFRGS, and the hospitality of the IFM-UFPel where this work was accomplished. This work was partially financed by the Brazilian funding agencies CNPq and FAPERGS.

\newpage

\begin{table}[t]
\begin{center}
\begin{tabular} {|l|l|l|}
\hline
 $\mathbf{\alpha_s}$  & \rm{{\bf scale}} & $\mathbf{\sigma_{tot}}$  \\
\hline
 $\alpha_s=0.300$ & $\simeq m^2_c$   & 4.88 pb \\
\hline
 $\alpha_s=0.289$ & $\simeq m^2_{J/\Psi}$   & 4.21 pb \\
\hline
 $\alpha_s=0.250$ & HERA   & 2.35 pb \\
\hline
 $\alpha_s=0.200$ & HERA   & 0.97 pb \\
\hline
\end{tabular}
\end{center}
\caption{ The estimates for $\sigma_{tot}(\gamma \gamma
\rightarrow J/\Psi \, J/\Psi)$ at the Born level for different
$\alpha_s$ values.} \label{tabela}
\end{table}

\newpage

\newpage

\newpage

\end{document}